\documentclass[aps,prb,twocolumn,superscriptaddress]{revtex4}
\usepackage[dvips]{epsfig}
\begin{document}

%\preprint{Ver 1}

\title{Optical diffraction spectroscopy of excitons in uniaxially-strained GaN films}


\author{Y. Toda}
\affiliation{Department of Applied Physics, Hokkaido University, Kita 13 Nishi 8 Kita-ku, Sapporo 060-8628, Japan.}
\affiliation{PRESTO Japan Science and Technology Agency, 4-1-8 Honcho, Kawaguchi, Saitama, Japan.}
\author{S. Adachi}
\affiliation{Department of Applied Physics, Hokkaido University, Kita 13 Nishi 8 Kita-ku, Sapporo 060-8628, Japan.}
\author{Y. Abe}
\affiliation{Department of Applied Physics, Hokkaido University, Kita 13 Nishi 8 Kita-ku, Sapporo 060-8628, Japan.}
\author{K. Hoshino}
\affiliation{Research Center for Advanced Science and Technology, Institute of Industrial Science, University of Tokyo, 4-6-1 Komaba, Meguro-ku, Tokyo, Japan.}
\author{Y. Arakawa}
\affiliation{Research Center for Advanced Science and Technology, Institute of Industrial Science, University of Tokyo, 4-6-1 Komaba, Meguro-ku, Tokyo, Japan.}

\date{\today}

\begin{abstract}
The degenerate four-wave mixing spectroscopy of uniaxially strained GaN layers is demonstrated using colinearly polarized laser pulses. The nonlinear response of FWM signal on exciton oscillator strength enhances the sensitivity for polarized exciton, allowing for mapping out the in-plane anisotropy of the strain field. The observed high-contrast spectral polarization clearly shows fine structure splittings of excitons, which are also confirmed in the change of quantum beating periods of time.
\end{abstract}

%\keywords{}

\maketitle
Influence of strain-fields on the optical properties of semiconductors has been studied extensively for many years. In particular, the presence of uniaxial strain generates the polarization and splitting of excitons, which provide useful information on exciton fine structures (EFSs) \cite{akimoto, Langer, Volm, Cho}. However, it is only recently that the EFS in wurtzite GaN was observed experimentally, where the exchange splitting was measured by M. Julier {\it et al.} as 0.6$\pm$0.1 meV using reflectivity measurements on a uniaxially strained GaN epilayer \cite{Julier}, and by P. Paskov {\it et al.} as 0.58$\pm$0.05 meV using photoluminescence (PL) measurements on thick GaN \cite{Paskov1,Paskov2}. The exchange interaction constant of $\gamma$ evaluated by a Hamiltonian based on quasi-cubic approximation shows a good correspondence in both measurements and is consistent with the exponential dependence on interatomic distance for III-V compounds. However, since the EFS is usually comparable to the exciton linewidth, more precise measurements are needed for detailed analysis. The use of nonlinear spectroscopy often provides significant enhancement of the signal sensitivity, which is advantageous for investigating a small amount of spectral changes such as those of EFSs.

In this study, we report on the polarization-resolved mapping of exciton lines in a uniaxially strained GaN layer using degenerate four-wave mixing (FWM) spectroscopy. Since the diffraction signal in FWM increases as the fourth power of the exciton oscillator strength, the exciton spectra show strong polarizations resulting from the uniaxial strain in the sample. The highly polarized spectra allow us to determine the precise splitting energy between excitons, which shows a good agreement with that obtained from the oscillation period of quantum beats.

The two-pulse FWM experiments in reflection geometry were performed using a frequency doubled, mode-locked tunable Ti:sapphire laser with the spectral width of 13 meV (FWHM) \cite{Adachi}. The excitation pulses with the same intensities were colinearly polarized and superimposed onto a sample surface using a lens (f= 200 mm). The FWM signal in the 2{\bf $k_2$-$k_1$} direction was spatially selected by an iris, and was spectrally resolved by a 30-cm-long monochromator with a CCD detector. The delay time $\tau_{12}$ between the pulses is defined as positive when the {\bf $k_1$} pulse precedes the {\bf $k_2$} pulse. The sample is mounted in a closed-cycle helium cryostat and all measurements were carried out at 10 K.

To understand the enhancement of sensitivity for exciton polarizations, we consider the FWM signal in a two-level system using the phase-space filling (PSF) nonlinearity \cite{Hazu}. For simplicity, we neglect additional many-body Coulomb interactions. Within the framework of the optical Bloch equation for the $\delta$-function pulses with the linear polarizations $E_{k_2}$ and $E_{k_1}$, the third-order optical polarization $P^{(3)}_{2k_2-k_1}$ at $\tau_{12}$ is described by \cite{Shah}
\begin{eqnarray*}
P^{(3)}_{2k_2-k_1} \propto -\frac{i}{2}\Theta(t-\tau_{12})\Theta(\tau_{12})E^2_{k_2}E^*_{k_1}|\mu_q|^2\mu_q, 
\end{eqnarray*}
where $\mu_q$ is the dipole matrix elements of linearly polarized excitons. As a consequence, the FWM signal $I_{FWM}$ with respect to the oscillator strength $|\mu_q|^2$ is expressed by
\begin{eqnarray*}
I_{FWM}(\tau_{12}) \propto \int dt |P^{(3)}_{2k_2-k_1}\cdot\mu_q^*|^2,
\end{eqnarray*}
indicating the fourth power dependence of $|\mu_q|^2$. The FWM signals thus provide a sensitive probe for linearly polarized excitons.

The sample used in this study is a high-quality 2.3 $\mu$m-thick GaN layer on an (11-20) a-plane sapphire substrate grown by metal-organic chemical vapor deposition \cite{Amano}. The details of the sample growth conditions were presented in Ref. 11.
In the sample, built-in uniaxial strain exists due to the anisotropic thermal expansion coefficients of the a-plane sapphire. Figure 1(a) shows the experimental geometry with respect to the crystallographic axis of the sample. The polarization angle $\theta$ of the incident pulses, which can be changed by rotating the half-wave plate (HWP) in front of the focusing lens, is measured from the [11-20] axis parallel to the direction of the uniaxial stress. Figure 1(b) illustrates an overview of the optically allowed states of A-exciton (X$^A$) and B-exciton (X$^B$) when the laser light is polarized perpendicular to the growth axis. In normal wurtzite GaN, there exists twofold degenerate spin-singlet states. The reduction of symmetry due to the uniaxial strain lifts the degeneracy and polarizes the optical transitions allowed in $x$ and $y$, where $x$ and $y$ are along the [11-20] and [1-100] axes, respectively.

\begin{figure}[ht]
\begin{center}
\includegraphics[width=85mm]{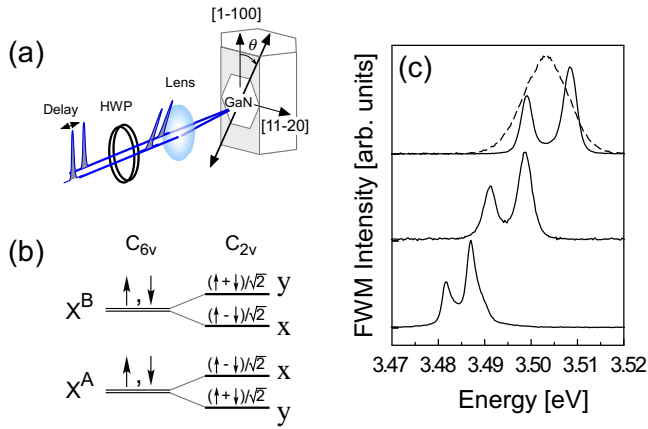}
\caption{(a) Experimental geometry. The colinear polarization angle $\theta$ of the incident pulses is defined from the [1-100] axis of GaN. (b) Schematic illustration of $\Gamma_5$ exciton levels in wurtzite GaN. Uniaxial in-plane strain reduces the symmetry from $C_{6v}$ (hexagonal) to $C_{2v}$ (orthorhombic). (c) Typical FWM spectra of GaN layer on a-plane sapphire (top), GaN layer on c-plane sapphire (middle), and free-standing GaN (bottom), at $\tau_{12}$ = 0 s. In the top spectra, the polarization angle $\theta$ is set to be $\pi$/4, and the dashed line indicates the excitation spectrum.}
\end{center}
\end{figure}
\begin{table}[hb]
\caption{\label{tab:table1}Exciton energies obtained from FWM spectra at 10 K.}
\begin{ruledtabular}
\begin{tabular}{llll}
Sample & X$^A$ (eV) & X$^B$ (eV) & X$^B-$X$^A$(meV)\\
\hline
Bulk GaN & 3.4816 & 3.4871 & 5.5\\
GaN on c-sapphire & 3.4912 & 3.4987 & 7.5\\
GaN on a-sapphire & 3.4990 & 3.5086 & 9.6\\
\end{tabular}
\end{ruledtabular}
\end{table}

We first consider the biaxial strain in order to determine several important parameters of crystal-field splitting $\Delta_{cr}$ and spin-orbit splitting $\Delta_{so}$. Figure 1 (c) shows typical FWM spectra at $\tau_{12}$ = 0. For comparison, FWM spectra of a 70 $\mu$m-thick free-standing GaN and a 2.3 $\mu$m-thick GaN layer on a (0001) c-plane sapphire are displayed in the figure. In the spectra, the lower- and upper-energy peaks are X$^A$ and X$^B$, respectively, each of which shows the systematic energy shifts depending on the sample. The exciton energies and their separations obtained from the Lorentzian fitting to the spectrum in each sample are listed in Table I. As has been reported by several groups \cite{Shikanai, Alemu}, the observed shifts reflect the biaxial strain associated with the mismatch of the lattice constants and of the thermal expansion coefficients between GaN and substrates in each sample. Following the analysis of A. Shikanai {\it et. al} \cite{Shikanai}, we obtained $\Delta_{cr}$ = 30 meV and $\Delta_{so}$ = 13 meV. 
\begin{figure}[ht]
\begin{center}
\includegraphics[width=75mm]{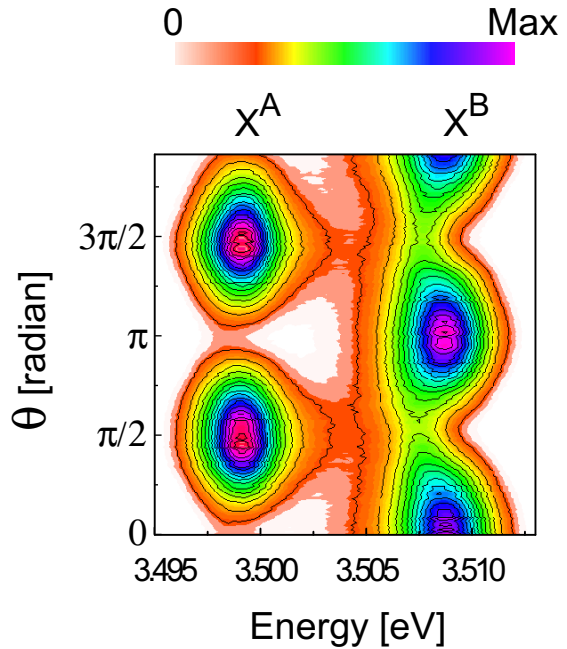}
\caption{(color). Contour plot of the FWM spectra for a GaN on a-plane sapphire taken by rotating the polarizaion $\theta$ at $\tau_{12}$ = 0. The spectra consists of peaks centered at approximately 3508.6 meV (X$^A$) and 3499.0 meV (X$^B$). The polarization angle of incident pulses was rotated in 1-degree steps and the integration time for collecting the signal at each $\theta$ was 1 sec.}
\end{center}
\end{figure}

The ability to map out the uniaxial strain-field is highlighted in Fig. 2, where we show a contour-plot of FWM spectra consisting of X$^A$ and X$^B$, recorded at the polarization angle of the incident pulses with a 1-degree step. The figure shows a striking feature of the anti-correlated polarizations in each exciton line, signifying the reduced symmetry due to the uniaxial strain. This optical anisotropy has been reported previously using the reflectance and the PL measurements \cite{Julier,Paskov2}. Note that compared with the case of linear spectroscopy, there is a considerable magnitude of polarization in which the nonlinear response in the FWM measurement increases by a factor of $\sim$10, allowing us to determine the uniaxial strain-field without fitting. Besides the nonlinearly enhanced exciton polarization, the contour-plot clearly shows the energy shifts originating from the EFS of two linearly polarized excitons. The lower-energy component of X$^A$ and the upper-energy component of X$^B$ are active at $\theta$ = 0, $\pi$ corresponding to the [1-100] axis (y-direction in Fig. 1 (b)), whereas the upper-energy component of X$^A$ and the lower-energy component of X$^B$ are active at $\theta$ = $\pi$/2, 3$\pi$/2 corresponding to the [11-20] axis (x in Fig. 1 (b)).
Thus, the  polarization spectra provides a type of crystalline analysis achieved in X-ray diffraction spectroscopy. Contrary to the X-ray diffraction, the FWM analyzes the sample using safe and simple experimental apparatus. We stress that this technique can be applied extensively to anisotropically polarized exciton systems with uniaxial strain.
\begin{figure}[t]
\begin{center}
\includegraphics[width=80mm]{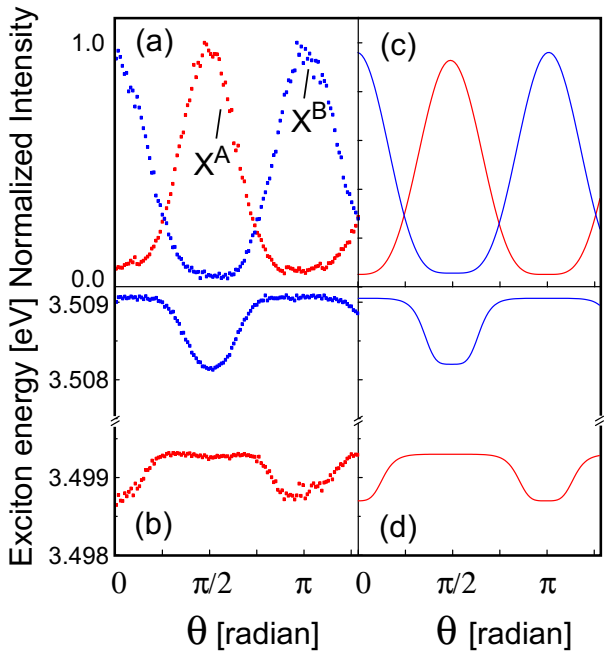}
\caption{Plots of (a) normalized peak intensities and (b) energies on A-exciton (red) and B-exciton (blue) lines obtained from the Lorentzian fitting to the FWM spectra at $\tau_{12}$ = 0.2 ps for various polarizations. Calculated polarization dependences of (c) FWM intensities and (d) peak energies for each exciton line.}
\end{center}
\end{figure}
\begin{table}[h]
\caption{\label{tab:table2}Fitting parameters for polarized FWM. E$^\alpha$ and $\Gamma^\alpha$ ($\alpha$=A,B) are exciton transition energy and linewidth (FWHM), respectively.}
\begin{ruledtabular}
\begin{tabular}{lllcll}
Polarization & E$^A$ (eV) & $\Gamma^A$ (meV) && E$^B$ (eV) & $\Gamma^B$ (meV)\\
\hline
X & 3.4993 & 2.1 && 3.5082 & 1.8\\
Y & 3.4987 & 2.3 && 3.5091 & 2.5\\
\end{tabular}
\end{ruledtabular}
\end{table}

For qualitative analysis with great precision, we made to fit the spectra with a combination of two Lorentzian functions. A Lorentzian lineshape is predicted in the case of a homogeneously broadened system, which will be confirmed in the time-evolutions of the signal. Figure 3 (a) and (b) plot the peak intensities and the energies for each exciton line at various $\theta$. Note that we evaluate the data obtained at $\tau_{12}\approx 0.2$ ps in order to reduce the contributions from efficient many-body effects at $\tau_{12}\approx 0$. In Fig. 3 (a), the variations of the FWM intensities deviate from the simple sinusoidal functions but agree well with the fourth power of them [$\sin^4(\theta)$], providing proof that the PSF nonlinearity accounts for our detection. The best fit to the data presented in Fig. 3 (c) gives the ratios of polarized FWM intensities for X$^A$ and X$^B$ as $I^{Ax}/I^{Ay}$ = 19.0$\approx$(2.1)$^4$ and $I^{By}/I^{Bx}$ = 17.2$\approx$(2.0)$^4$, respectively. This results in a trapezoidal variation of the spectral peak energies (Fig. 3(b)). Based on the ratios of polarized FWM intensities, we can reproduce such polarization dependences as shown in Fig. 3 (d). The parameters obtained from these analysis are listed in Table II. The splitting energies between the colinearly polarized EFS in each exciton are E$^B$$-$E$^A$ = 8.9 meV (x) and 10.4 meV (y). The average energy shift of the EFS is thus $\sim$0.8 meV. Using the specific parameters obtained from the biaxial strain, we are now ready to evaluate the exchange interaction constant $\gamma$ for this sample. Following the analysis presented by M. Julier {\it et al}. \cite{Julier}, we obtain $\gamma$ = 0.6 meV using the uniaxial strain energy $\delta_3$ = $-$1.5 meV. 
\begin{figure}[hb]
\begin{center}
\includegraphics[width=75mm]{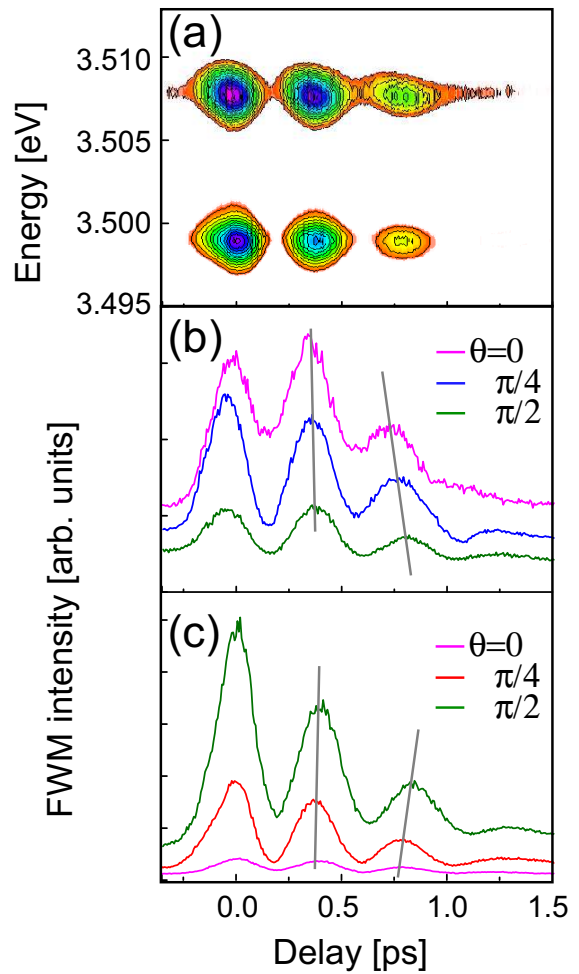}
\caption{(a) Contour plot of FWM spectra at $\theta$ = $\pi$/4 for various $\tau_{12}$. FWM intensities for (b) X$_B$ and (c) X$_A$ as functions of delay $\tau_{12}$ at different incident polarizations $\theta$. The gray lines are guides for the eye, indicating the changes in quantum beating oscillation $\Omega$ for various $\theta$.}
\end{center}
\end{figure}

In addition to the spectral evolutions, FWM experiments allow us to examine the temporal dynamics of excitons in EFS. Figure 4 (a) shows a contour plot of FWM as a function of $\tau_{12}$, at $\theta$ =$\pi$/4. As is well known, the simultaneously excited two excitons result in a quantum beating oscillation, of which the oscillation period $\Omega$ corresponds to the energy separation between excitons. In Figs. 4 (b) and (c), we plot the peak intensities of each resonance at $\theta = 0, \pi/4, \pi/2$. The polarization dependence of $\Omega^{A,B}(\theta)$ in each polarization configuration is clearly seen, and is, of course, comparable in each exciton line ($\Omega^{A}(\theta) \approx \Omega^{B}(\theta)$).
Assuming the homogeneously broadened excitons, the FWM signal at $\tau_{12}$ can be expressed as 
\begin{eqnarray*}
&&I^{A,B}_{FWM}(\tau_{12}, \theta) \propto \\
&&[1 + a^{A,B}(\theta) \cos(\Omega(\theta)\tau_{12} - \phi^{A,B})]\exp(-2\tau_{12}/T_2^{A,B}(\theta)).
\end{eqnarray*}
In this fitting function, $a^{A,B}$ denotes the beating contrast and $T_2^{A,B}$ denotes the dephasing time, both of which depend on $\theta$. Neglecting the many-body Coulomb interactions, we consider the quantum beats dominated by the two pairs of colinearly polarized excitons, X$^{Ax}$-X$^{Bx}$ and X$^{Ay}$-X$^{By}$. In this assumption, the results obtained at $\theta$ = 0 and $\pi$/2 reflect the natures of X$^x$ and X$^y$, respectively. The exact phase $\phi^{A,B}$ is slightly different in each polarization mainly due to the Coulomb interactions between excitons \cite{Aoki}. For simplicity, we also neglect these contributions. From the fitting to the data, we obtain $\Omega^x$ = 13.6 rad./ps and $\Omega^y$ = 15.7 rad./ps. These parameters correspond to the splitting energies between the colinearly-polarized EFS in each exciton of E$^B$-E$^A$ = 8.95 meV (x) and 10.3 meV (y), coinciding with those obtained from the polarization-resolved spectra.

The dephasing times in each exciton are $T_2^{Ax}$ = 0.58 ps, $T_2^{Ay}$ = 0.60 ps, $T_2^{Bx}$ = 0.70 ps, and $T_2^{By}$ = 0.58 ps, which are comparable with those obtained from the Lorentzian fits to the FWM spectra. Thus, we conclude that exciton lines are homogeneously broadened in our sample.

Also, the beating contrasts $a^{A,B}(\theta)$ are $a^A(0)\approx a^B(0) \approx 0.6 $ and $a^A(\pi/2)\approx a^B(\pi/2) \approx 0.9$. The polarization dependence is mainly due to the difference in relative oscillator strength between the excitons. On the basis of a noninteracting three-level system, the beating contrast is given by\cite{Leo}
\begin{eqnarray*}
a^{A,B}(\theta) = \frac{2|\mu_A(\theta)|^2|\mu_B(\theta)|^2}{|\mu_A(\theta)|^4+|\mu_B(\theta)|^4}.
\end{eqnarray*}
According to the spectral evolutions, we assume the relative exciton oscillator strengths to be $|\mu_A(0)|^2$ = 0.31, $|\mu_B(0)|^2$ = 1.0, $|\mu_A(\pi/2)|^2$ = 0.65, and $|\mu_B(\pi/2)|^2$ =0.49. As a result, we obtain the beating contrasts as $a^{A,B}(0) \approx 0.57$ and $a^{A,B}(\pi/2) \approx 0.96$, which give excellent agreement with those evaluated using the time evolutions.

Finally, We comment on the splitting of EFS in each exciton. In Fig. 3(b), the shifts of EFS energy are $E^{Ax}-E^{Ay}$ = 0.6 meV and $E^{By}-E^{Bx}$ = 0.9 meV, showing a difference in splitting energy between X$^A$ and X$^B$. This difference between excitons is consistently observed throughout all our FWM experiments, and becomes even more apparent in Fig. 2, where, in the absence of any fitting, one can see a clear difference in the distributions of each exciton line. Furthermore, we have obtained a similar difference in the reflectivity measurements, excluding the contributions of many-body effects. Therefore, we conclude that the different splittings originate from the difference in EFS between excitons of our sample, suggesting the breakdown of the quasi-qubic approximation in wurtzite GaN \cite{Thang}. Further investigations including theoretical supports are needed to clarify the difference. However, we emphasize the importance of the enhanced sensitivity for exciton polarizations achieved in this measurement that allows us to account for the small amount of spectral changes in EFSs.

In conclusion, a combination of the fourth power dependence of exciton oscillator strength on a four-wave mixing signal and the built-in uniaxial strain of the sample allow not only for a type of crystalline analysis by X-ray diffraction spectroscopy, but also for studying exciton fine structures and their temporal dynamics. This technique can be applied to optical characterization for polarized exciton systems with an appropriate anisotropic strain field.

The authors acknowledge Dr. T. Someya for useful discussions. This work is partly supported by Grant-in-Aid for Scientific Research and IT program, MEXT. K. H. acknowledges Research Fellowship of the JSPS for Young Scientists.

%\end{acknowledgments}

\end{document}